\documentclass[sigconf, authorversion]{acmart}

\usepackage{balance} 
\usepackage{multirow} 
\usepackage{hyperref} 
\usepackage{enumitem} 

\usepackage{todonotes}
\usepackage{fvextra} 
\usepackage{tcolorbox}
\usepackage{tabularx}

\AtBeginDocument{%
  \providecommand\BibTeX{{%
    \normalfont B\kern-0.5em{\scshape i\kern-0.25em b}\kern-0.8em\TeX}}}

\copyrightyear{2023}
\acmYear{2023}
\setcopyright{rightsretained}
\acmConference{CACM}{Accepted: June}{2023}

\begin{document}

\title[Computing Education in the Era of Generative AI]{Computing Education in the Era of Generative AI}

\author{Paul Denny}
\orcid{0000-0002-5150-9806}
\affiliation{
  \institution{The University of Auckland}
  \city{Auckland}
  \country{New Zealand}
}
\email{paul@cs.auckland.ac.nz}

\author{James Prather}
\orcid{0000-0003-2807-6042}
\affiliation{
  \institution{Abilene Christian University}
  \city{Abilene, Texas}
  \country{USA}
}
\email{james.prather@acu.edu}

\author{Brett A. Becker}
\orcid{0000-0003-1446-647X}
\affiliation{
  \institution{University College Dublin}
  \city{Dublin}
  \country{Ireland}
}
\email{brett.becker@ucd.ie}

\author{James Finnie-Ansley}
\orcid{0000-0002-4279-6284}
\affiliation{
  \institution{The University of Auckland}
  \city{Auckland}
  \country{New Zealand}
}
\email{james.finnie-ansley@auckland.ac.nz}

\author{Arto Hellas}
\orcid{0000-0001-6502-209X}
\affiliation{
  \institution{Aalto University}
  \city{Espoo}
  \country{Finland}
}
\email{arto.hellas@aalto.fi}

\author{Juho Leinonen}
\orcid{0000-0001-6829-9449}
\affiliation{
  \institution{Aalto University} 
  \city{Espoo}
  \country{Finland}
}
\email{juho.2.leinonen@aalto.fi}

\author{Andrew Luxton-Reilly}
\orcid{0000-0001-8269-2909}
\affiliation{
  \institution{The University of Auckland}
  \city{Auckland}
  \country{New Zealand}
}
\email{a.luxton-reilly@auckland.ac.nz}

\author{Brent N. Reeves}
\email{brent.reeves@acu.edu}
\orcid{0000-0001-5781-1136}
\affiliation{%
  \institution{Abilene Christian University}
  \city{Abilene, Texas}
  \country{USA}
}

\author{Eddie Antonio Santos}
\orcid{0000-0001-5337-715X}
\affiliation{
  \institution{University College Dublin}
  \city{Dublin}
  \country{Ireland}
}
\email{eddie.santos@ucdconnect.ie}

\author{Sami Sarsa}
\orcid{0000-0002-7277-9282}
\affiliation{
  \institution{Aalto University}
  \city{Espoo}
  \country{Finland}
}
\email{sami.sarsa@aalto.fi}

\renewcommand{\shortauthors}{Denny et al.}

\begin{abstract}

The computing education community has a rich history of pedagogical innovation designed to support students in introductory courses, 
and to support teachers in facilitating student learning.  Very recent advances in artificial intelligence have resulted in code generation models that can produce source code from natural language problem descriptions --- with impressive accuracy in many cases. The wide availability of these models and their ease of use has raised concerns about potential impacts on many aspects of society, including the future of computing education. In this paper, we discuss the challenges and opportunities such models present to computing educators, with a focus on introductory programming classrooms. 
We summarize the results of two recent articles, the first evaluating the performance of code generation models on typical introductory-level programming problems, and the second exploring the quality and novelty of learning resources generated by these models. We consider likely impacts of such models upon pedagogical practice in the context of the most recent advances at the time of writing. 

\end{abstract}



\maketitle

\section{Introduction}


A new era is emerging in which artificial intelligence will play an ever-increasing role in many facets of daily life.  
One defining characteristic of this new era is the ease of generating novel content, including natural language, with remarkable speed and quality.
Large language models (LLMs), which are a type of foundation model \cite{bommasani2021opportunites}, are neural network-based language models that are trained on vast quantities of text data.
Such models are capable of creating a variety of convincing human-like outputs including prose, poetry, and source code.  It is largely accepted that synthesizing source code automatically from natural language descriptions of problems is likely to greatly improve the productivity of professional developers~\cite{peng2023impact},
and is being actively explored by well-funded entities such as OpenAI (ChatGPT, Codex, GPT-\textit{x}), Amazon (CodeWhisperer), and Google (AlphaCode, Bard).  In the same way that high-level programming languages offered large productivity advantages over assembly language programming in the 1970s, AI code generation tools look set to revolutionize traditional programming in the coming years. Already, claims are emerging that a significant proportion of new code is being produced by tools such as GitHub Copilot\footnote{\href{https://github.blog/2023-03-22-github-copilot-x-the-ai-powered-developer-experience/}{https://github.blog/2023-03-22-github-copilot-x-the-ai-powered-developer-experience/}}, 
and the current pace of development in this area is staggering. Noticeably more advanced model versions are being released several times per year, despite the first only being made generally available less than two years ago. The pace of advancement is so rapid that in March 2023, a well-publicized open letter appeared that encouraged a public, verifiable, and immediate pause of at least six months duration on the training of AI systems more powerful than GPT-4. Signed by Elon Musk, Steve Wozniak, Moshe Vardi and thousands of others including many AI-leaders and Turing award winners\footnote{\href{https://futureoflife.org/open-letter/pause-giant-ai-experiments/}{futureoflife.org/open-letter/pause-giant-ai-experiments/}}, the letter was addressed to all AI labs and suggested potential government-led moratoriums. 



These developments raise urgent questions about the future direction of society, including education, and specifically computing education. In particular, the ease with which students can auto-generate solutions to programming assessments focuses new attention on the well-studied area of introductory programming~\cite{becker201950,luxton-reilly2018introductory}.
One popular evidence-based pedagogy involves students gaining extensive practice writing code through the use of many small exercises~\cite{vihavainen2011extreme}
that are checked either manually or by automated grading tools. 
However, many such problems can now easily be solved by AI models. Often all that is required of a student is to accept an auto-generated suggestion by an IDE plugin~\cite{finnieansley2022robots}.   This casts doubt on the value of current pedagogical approaches, and raises concerns that students may use new tools in ways that limit learning and that make the work of educators more difficult.   For example, in a deep discussion of LLMs, Bommasani et al. highlight that it will become ``much more complex for teachers to understand the extent of a student's contribution'' and to ``regulate ineffective collaborations and detect plagiarism''~\cite{bommasani2021opportunites}.

Alongside such challenges come emerging opportunities for students to learn computing skills~\cite{becker2023programming}. Learners without access to formal education opportunities can produce code 
directly within their IDEs (e.g. using GitHub Copilot) and can seek help at any time when faced with difficulties using dialogue-based models (e.g. ChatGPT). 
These tools are already being used in the classroom by both teachers and students.  
 
In this article, we consider the implications of generative AI on computing education, and explore how newly emerging tools are likely to impact students and educators in introductory programming classrooms in the near future.  We organize the article into two main sections: challenges and opportunities.  With respect to challenges, we summarize the results of recent work evaluating the accuracy and performance of code generation models on typical introductory level programming problems.  We also discuss issues relating to plagiarism, learner over-reliance and potential risks around bias and bad habits. With respect to opportunities, we summarize the results of recent work exploring how these models can be used to generate learning resources, including programming exercises and code explanations.  We further discuss the potential for improving feedback to students, such as error message reporting, rapid solutioning, and new pedagogical approaches.

\subsection{Large Language Models and Code }
\label{sec:details}
Although long-sought, AI-driven code generation has only been a viable reality for the general public since 2021. Several large-scale AI-powered code generation tools have recently been released including OpenAI's ChatGPT, DeepMind's AlphaCode, and Amazon's CodeWhisperer. 
These systems claim to make ``programming more productive and accessible''~\cite{li2022alphacode}, and to be ``Your AI pair programmer''\footnote{\href{https://github.com/features/copilot}{github.com/features/copilot}}. Codex (discussed in this article specifically) is a descendant of GPT-3, an autoregressive language model capable of producing text similar to that produced by humans. As a version of GPT-3, Codex is fine-tuned with code from over 50 million public GitHub repositories totaling 159\,GB~\cite{chen2021evaluating}. The model can take English-language prompts and generate code in several programming languages including JavaScript, Go, Perl, PHP, Ruby, Swift, TypeScript, and shell, but is ``most capable'' in Python\footnote{\href{https://doi.org/10.48550/arXiv.2211.03622}{doi.org/10.48550/arXiv.2211.03622}}. 
It can also translate code between programming languages, explain (in several natural languages) the functionality of code provided as input, and return the time complexity of code it generates.  Despite a claim that it would be deprecated in March 2023, Codex is still available for researchers via the OpenAI API\footnote{\href{https://beta.openai.com/}{beta.openai.com}}, and an upgraded version still powers GitHub Copilot\footnote{\href{https://copilot.github.com/}{copilot.github.com}}, a plug-in for popular IDEs such as Visual Studio Code.  However, newer tools, such as GPT-4, are often better at code-related tasks and should be utilized for future research. 

The use of such tools in education is nascent. Copilot was only made freely available to students in June 2022\footnote{\href{https://github.blog/2022-06-21-github-copilot-is-generally-available-to-all-developers/}{github.blog/2022-06-21-github-copilot-is-generally-available-to-all-developers}}, and to teachers in September 2022, after the potential to impact education began to unfold.\footnote{\href{https://github.blog/2022-09-08-github-copilot-now-available-for-teachers/}{github.blog/2022-09-08-github-copilot-now-available-for-teachers}} In November 2022, ChatGPT\footnote{\href{https://openai.com/blog/chatgpt/}{openai.com/blog/chatgpt}} was released. It presents users with a conversational interface which maintains contextual information.  ChatGPT garnered wide publicity for the quality of its outputs. As one example, it has alarmed educators who rely on written essays for assessment, as ChatGPT appears capable of producing written essays of surprisingly good quality on almost any topic except recent events. It can also seamlessly interweave code and text. 
In March 2023, GPT-4 was released. As of the writing of this article, it is available via Open AI's Chat-GPT Plus and as an API for developers and researchers with access via a waitlist.  

It appears certain that the adoption of such tools by both students and teachers will increase in tandem with improvements in model capabilities.  For a more technical overview of the historical developments and future trends of language models, we direct the reader to the recent \emph{CACM} article by Li~\cite{li2022language}.

\section{Challenges ahead}
Code generation tools powered by LLMs can correctly and reliably solve many typical introductory programming (often called CS1~\cite{becker201950}) problems. This raises interesting questions for educators including: Just how good are these tools? Can students with no programming knowledge, but who are armed with these tools, pass typical programming-centric assessments that are common in introductory courses?  Do we need a different approach?

%
%
\subsection{Putting Them to the Test}

To explore the performance of LLMs in the context of introductory programming, we prompted Codex with real exam questions and compared its performance to that of students taking the same exams~\cite{finnieansley2022robots}. We also prompted Codex to solve several variants of a well-known CS1-level programming problem (the ``Rainfall problem'') and examined both the correctness and the variety of solutions produced.

\subsubsection{My AI wants to know if its grade will be rounded up.}
We took all questions from two Python CS1 programming exams that had already been taken by students and provided them as input  (verbatim) to Codex. The exam questions involved common Python datatypes including strings, tuples, lists, and dictionaries.  They ranged in complexity from simple calculations, such as computing the sum of a series of simulated dice rolls, to more complex data manipulations such as extracting a sorted list of the keys that are mapped to the maximum value in a dictionary.

To evaluate the code generated, we executed it against the same set of test cases that were used in assessing the student exams. This follows a similar evaluation approach employed by the Codex developers~\cite{chen2021evaluating}. If the Codex output differed from the expected output with only a trivial formatting error (for example, a missing comma or period) we made the appropriate correction, much as a student would if using Codex to complete an exam.

\begin{figure}[]
    \centering
    \includegraphics[width=.95\linewidth]{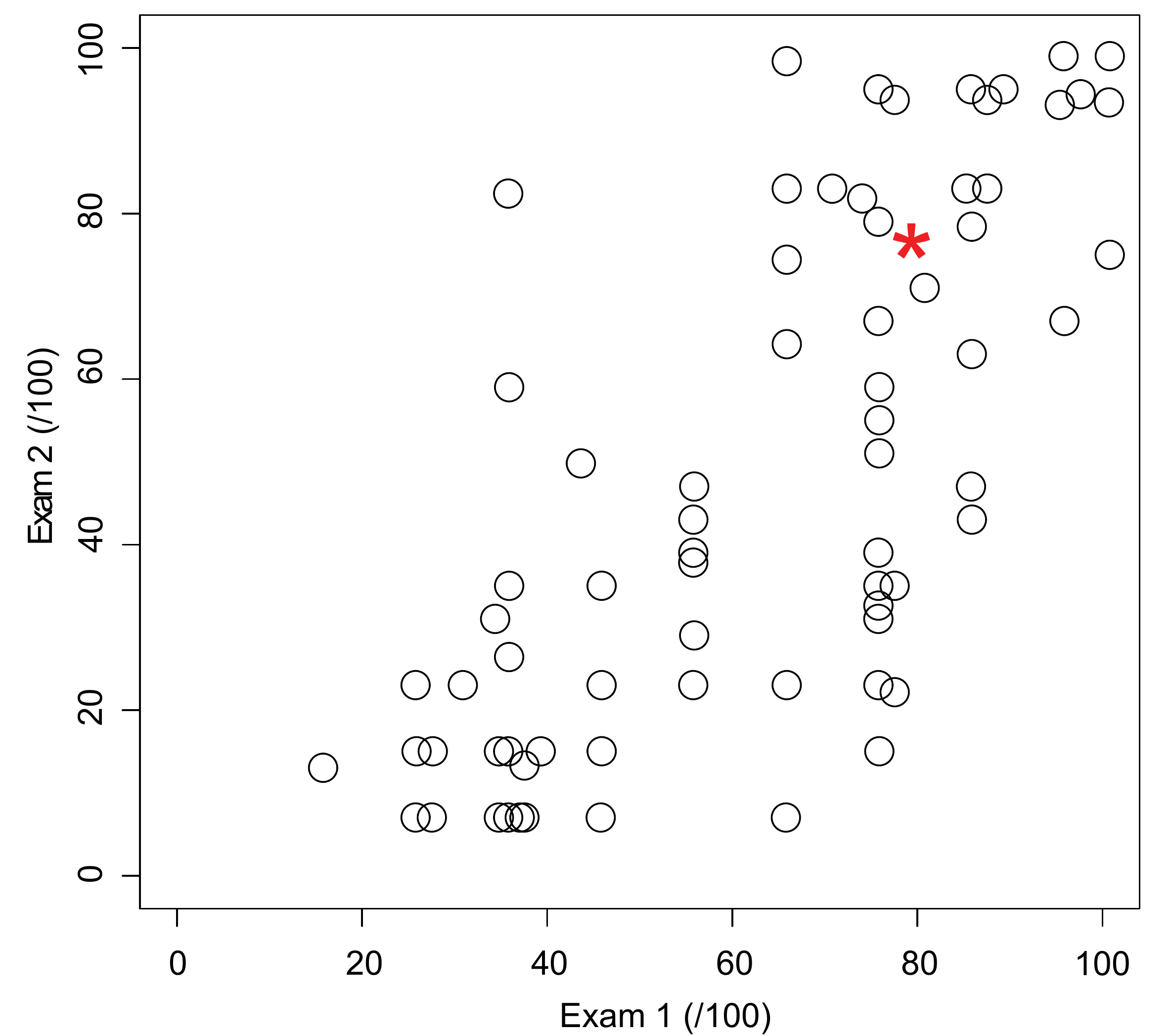}
    \caption{Student scores on Exam 1 and Exam 2. Codex's score is represented by the red asterisk.}
    \label{fig:class-scores}
\end{figure}

To contextualize the performance of the Codex model, we calculated the score for its responses in the same way as for real students using the same question weights and accumulated penalties for incorrect submissions. Codex  scored 15.7/20 (78.5\%) on Exam 1 and 19.5/25 (78.0\%) on Exam 2.  Figure~\ref{fig:class-scores} plots the scores (scaled to a maximum of 100) of 71 students enrolled in the CS1 course in 2020 who completed both exams. Codex's score is marked with a red asterisk. Averaging both Exam 1 and Exam 2 performance, Codex ranks 17 amongst the 71 students, placing it within the top quartile of class performance.

We observed that some of the Codex answers contained trivial formatting errors. We also observed that Codex performed poorly with problems that disallowed the use of certain language features (e.g. using \texttt{split()} to tokenize a string). Codex often did not produce code that avoided using these restricted features, and thus the model (in these cases) often did not pass the auto-grader. Codex also performed poorly when asked to produce formatted ASCII output such as patterns of characters forming geometric shapes, especially where the requirements were not specified in the problem description but had to be inferred from the provided example inputs and outputs.


\subsubsection
{Yes, I definitely wrote this code myself.}
\label{sec:yes}
To understand the amount of variation in the responses, we provided Codex with seven variants of the problem description for the well-studied Rainfall problem (which averages values in a collection) 
a total of 50 times each, generating 350 responses. Each response was executed against 10 test cases (a total of 3500 evaluations). Across all variants, Codex had an average score close to 50\%. 
Codex performed poorly on cases where no valid values were provided as input (e.g. where the collection to be averaged was empty).


We also examined the number of source lines of code for all Rainfall variants, excluding blank and comment lines. In addition, we classified the general algorithmic approach employed in the solutions as an indicator of algorithmic variation. We found that Codex provides a diverse range of responses to the same input prompt. Depending on the prompt, the resulting programs used varied programmatic structures, while ultimately favoring expected methods for each problem variation (i.e., for-loops for processing lists, and while-loops for processing standard input).

\subsection{Academic Integrity}
Software development often encourages code reuse and collaborative development practices, which makes the concept of academic integrity difficult to formalize in computing~\cite{simon.ea-2016-maze}. Nevertheless, individual work is still commonplace in computing courses, and it is an expectation for students working on individual projects to produce their own code rather than copying code written by someone else.  This is often verified through the use of plagiarism tools which analyze code structure and are very effective at detecting similarities between an original submission and one that is produced by copying and then modifying the original. However, recent work has shown that common plagiarism detection tools are often ineffective against AI-generated solutions~\cite{biderman2022fooling}. This raises significant concerns for educators monitoring academic integrity in formal assessments.

\subsubsection{Academic misconduct \ldots}
The observations above demonstrate that AI-generated code tools can provide students with the ability to perform better than average on exams with typical CS1-style problems~\cite{finnieansley2022robots}.
Other recent work has shown that these tools can reliably generate correct code for common algorithms such as insertion sort and tree traversal~\cite{dakhel2022github}, and for problems that are typical in more advanced computing courses focusing on data structures and algorithms \cite{finnieansley2023my}. Presently, it is reasonable for educators to assume that students are using these tools --- many already are.

One type of academic misconduct that has become more common in recent years is contract cheating --- the outsourcing of assignments to other humans ~\cite{simon.ea-2016-maze}. This is difficult to detect, but it is not without risks. Contract cheating opens the student to blackmail if the agreed price is not paid, or if the contractor decides to extort the student for more money (events which are, anecdotally, becoming more common). It is also possible that the contractor could reuse submissions, which would be flagged by traditional plagiarism detection software. On the other hand, AI-generated solutions that exhibit a great deal of variation~\cite{finnieansley2022robots} leave no paper trail of communication with another human, produce similar results to outsourcing, and incur minimal cost. For students who are time-poor and focused on the short-term reward of grades, this low-risk option may appear alarmingly attractive.

\subsubsection{\ldots or not?}
Simon et al. surveyed academics about the use of attribution for code obtained from outside sources, finding a diverse range of views on the acceptability of code reuse~\cite{simon.ea-2016-maze}.  This academic integrity quagmire becomes more complex with relatively opaque differences between standard code completion tools present in IDEs and plugins such as Copilot that provide code suggestions that may appear indistinguishable from IDE-based code completion.

The use of spelling- and grammar-checking, calculators, or IDE code completion tools does  not typically raise academic integrity concerns. Although these tools all perform tasks that once required humans, we have accepted that they are readily available, and their use does not cause concern in most instances. However, many of these innovations caused quite a stir when they were new. Most educators do not expect students to attribute a calculation (for instance) to a machine (such as a calculator), despite the machine generating the result. With the benefit of hindsight, will the present moment just be a ``calculator'' moment for a wider range of disciplines?  Should we be treating the use of machine-generated code as academic misconduct, and if so, how should we distinguish between acceptable and unacceptable machine-generated code suggestions? 
 
In the short term, we anticipate some institutional knee-jerk reactions --- perhaps more frequently using secure assessment conditions. It is possible that increased attention to data that is linked to individual activity such as keystroke data 
may be utilized by some. Regardless, the key philosophical issue remains: how much content (and what nature of content) can be machine-generated while still attributing the intellectual ownership to a human?  This calls into question the very concept of plagiarism~\cite{dehouche2021plagiarism} and how we should interpret plagiarism and intellectual contribution with machine-supported content generation. In fact, AI-generated code and license infringement are already heading for the courts (see Section~\ref{licensing}). We  must engage with these questions to ensure we have appropriate disciplinary perspectives on academic integrity in computing --- most importantly ones that are not orthogonal to the ultimate goal of learning.


%

\subsubsection{Code reuse and licensing}
\label{licensing}
Potential licensing issues arise when content is produced using code generation models, even when the model data is publicly-available~\cite{li2022alphacode}.  Many different licenses apply to much of the publicly-available code used to train LLMs, and typically these licenses require authors to credit the code they used, even when the code is open-source.  When a developer generates code using an AI model, they may end up using code that requires license compliance without being aware of it.  This controversy has led to a class-action lawsuit which claims that Copilot violates the rights of creators who shared their code under open source licenses\footnote{\href{https://githubcopilotlitigation.com/}{githubcopilotlitigation.com}}.  This is clearly an issue that extends beyond educational use of software, but as educators it is our role to inform students of their professional responsibilities when reusing code. 

The increasing use of AI-generated code provides ample opportunity for discussions around ethics and the use of computers in society. Moreover, these technologies may serve as a vehicle to empower novice users to 
explore more advanced ideas earlier,
leveraging the natural engagement that comes from utilizing technologies that are `in the news'. Teachers of introductory courses have long told themselves that students will learn about testing, security, and other more advanced topics in subsequent courses. However, with growing numbers of students taking introductory classes but not majoring in computing, and the capabilities that code generation affords, the stakes are higher for CS1 and introductory classes to raise these issues early, before the chance of real-world harm is great.

\subsection{Learner Over-reliance}

\subsubsection{Developing metacognition}
The developers of Codex noted that a key risk of using code generation models in practice is users' over-reliance~\cite{chen2021evaluating}. Novices using such models, especially with tools such as Copilot that embed support in an IDE, may quickly become accustomed to auto-suggested solutions.  This may lead to students not reading problem statements carefully, or at all, and therefore not thinking about the computational steps needed to solve a problem. 

Developing computational thinking skills is important for novice programmers as it can foster higher-order thinking and reflection skills~\cite{loksa2022metacognition}. 
Metacognition, or ``thinking about thinking'', is a key aspect of computational thinking (and problem solving in general) and has been shown to be closely related to it. 
While learning to code is already a challenging process that requires a high level of cognitive effort to remember language syntax, think computationally, and understand domain-specific knowledge, the use of metacognitive knowledge and strategies can aid in problem-solving and prevent beginners from getting overwhelmed or lost. 
These higher-order skills extend beyond individual self-regulation, and include helping other individuals (co-regulation), and groups (socially-shared regulation)~\cite{prather2022getting}. Relying too heavily on code generation tools may hinder the development of these crucial metacognitive skills.

\subsubsection{When the models fail}
An analysis of solutions generated by AlphaCode revealed that 11\% of Python solutions were syntactically-incorrect (produced a \texttt{SyntaxError}) and 35\% of C++ solutions did not compile~\cite{li2022alphacode}. Recent work has shown that as many as 20\% of introductory programming problems are not solved sufficiently by code generation models, even when allowing for expert modification of the natural language problem descriptions~\cite{denny2022conversing}. The developers of Codex noted that it can recommend syntactically-incorrect code including variables, functions, and attributes that are undefined or outside the scope of the codebase, stating ``Codex may suggest solutions that superficially appear correct but do not actually perform the task the user intended. This could particularly affect novice programmers, and could have significant safety implications depending on the context''~\cite{chen2021evaluating}. 
Students who have become overly reliant on model outputs may find it especially challenging to proceed when suggested code is incorrect and cannot be resolved through natural language prompting.



\subsection{Bias and Bad Habits}

The issue of bias in AI is well known
. In addition to general bias (subtle or overt) that applies to almost all AI-generated outputs, such as the representation of certain groups of people, genders, etc., there are likely biases specific to AI code generation.  

\subsubsection{Appropriateness for beginners}
Novices usually start by learning simple programming concepts and patterns and gradually building their skills.  However, much of the vast quantity of code these AI models are trained on is written by experienced developers, and is complex.  
AI generated code may be too advanced or complex for novices to understand and modify.  It may contain stylistic elements, or use approaches, that are intentionally not taught by an instructor. 

\subsubsection{Harmful biases}
The developers of Codex found that code generation models raise bias and representation issues 
--- notably that Codex can generate code comments (and potentially identifier names) that reflect negative stereotypes about gender and race, and may include other denigratory outputs~\cite{chen2021evaluating}. 
Such biases are obviously problematic, especially where novices are relying on the outputs for learning purposes.  

\subsubsection{Security}

Unsurprisingly, AI-generated code can be insecure~\cite{pearce2022asleep}, and human oversight is required for the safe use of AI code generation systems ~\cite{chen2021evaluating}.  However, novice programmers lack the knoweldge to provide this oversight.  Perry et al. recently examined whether novices using AI code generation tools wrote more secure code, finding that novices consistently wrote insecure code with specific high vulnerabilities in string encryption and SQL injection~\cite{perry2022users}.  Perhaps even more disturbing, novice programmers in their study who had access to an AI code generating tool were more likely to believe they had written secure code. 

Chen et al. noted that although future code generation models may be able to produce more secure code than the average developer, this is far from certain~\cite{chen2021evaluating}. CodeWhisperer claims to tackle security head-on by providing the ability to run scans on code 
to detect security vulnerabilities. 
Thus, there is a pressing need for increased student and educator awareness around the limitations of current models for generating secure code.


\section{Opportunities ahead}


Despite the challenges that must be navigated, code generation tools have the potential to revolutionize teaching and learning in the field of computing~\cite{becker2023}. 
Indeed, the developers of such models specifically highlight their potential to positively impact education.  When introducing Codex, Chen et al. outline a range of possible benefits, including to: ``aid in education and exploration''~\cite{chen2021evaluating}. Similarly, the developers of AlphaCode suggest such tools have ``the potential for a positive, transformative impact on society, with a wide range of applications including computer science education''~\cite{li2022alphacode}.   However, precisely how to realize this potential remains unclear.  In this section we discuss several concrete opportunities for code and text generation models to have a transformative effect on computing education.




%
%

\subsection{Plentiful Learning Resources}

Introductory programming courses typically utilize a wide variety of learning resources.  For example, programming exercises are a common type of resource that are presented together with automated assessment tools which students use to develop mastery through regular practice.  One of the key benefits of small programming exercises is that they impose low cognitive load, and thus improve the efficiency of learning targeted concepts. 
However, crafting good programming exercises with tests that verify solutions is a known challenge for educators. 
The cost to develop these exercises is exacerbated when students share their solutions.


Natural language explanations of code are another useful resource.  They can be valuable for helping students understand how a complex piece of code works, or as a tool for evaluating student learning.
The ability to reason about code and explain its purpose is a key skill that novices develop as they gain expertise, and `explain in plain English' questions are a common exercise used to scaffold this skill.
Generating explanations of code, both to help students when stuck and as exemplars for students to learn from, requires significant instructor effort. 
We explored the potential for LLMs to reduce the effort needed by instructors to 
generate the two types of learning resources just discussed: programming exercises and code explanations~\cite{sarsa2022automatic}.

\subsubsection{Programming exercises}

 Figure~\ref{fig:priming} shows an example of the input we used to generate new programming exercises using Codex. This `priming' exercise consists of a one-shot example (a complete example similar to the desired output) followed by a partial prompt to prime the generation of a new output.  In this case, the format of the priming exercise consists of a label (Exercise 1) followed by keywords for both the contextual themes (donuts) and the programming-related concepts (function, conditional) of the exercise, a natural language problem statement and a solution (in the form of a Python function). For space reasons, we omit a list of test cases but these can also be included for programming problems.  The priming input ends with the explicit prompt for a new exercise to be generated (Exercise 2), along with the desired concepts and themes expressed as keywords (basketball, function, list, and for loop). 

\begin{figure}[b]

\begin{tcolorbox}
\begin{Verbatim}[fontsize=\fontsize{9pt}{9pt},breaklines]
"""Exercise 1
--Keywords--
donuts
function
conditional
--Problem statement--
Write a function called donut_cost that takes three inputs: the price of a donut, the number of donuts bought, and whether or not it is the weekend.  The function should return the total cost of the donuts.  If it is the weekend, the donuts cost twice as much.
--Sample solution--
def donut_cost(price, number, is_weekend):
  if is_weekend: 
    return price * number * 2
  else: 
    return price * number
"""Exercise 2
--Keywords--
basketball
function
list
for loop
--Problem statement--
\end{Verbatim}
\end{tcolorbox}

\caption{A `priming' exercise consisting of one complete example followed by the prompt for a new one.}
\label{fig:priming}
\end{figure}

Figure~\ref{fig:new-exercise} shows one output generated when the prompt in Figure~\ref{fig:priming} was provided to Codex.  Note that in this case, as requested by the keyword information in the input prompt, the problem statement is related to basketball and the model solution consists of a \emph{function} that involves a \emph{list} and a \emph{for} loop.  To evaluate this approach more thoroughly, we generated a set of 240 programming exercises by varying the programming related concepts and contextual themes.  We attempted to execute the generated code against the generated test cases, and analyzed statement coverage as a measure of the thoroughness of the test suite.  Table \ref{tab:dynamic_analysis_results} summarizes these results, and shows that in most cases the programming exercise generated by the model included a sample solution that was executable.  Similarly, most of the time the model also generated a set of tests, resulting in a total of 165 programming exercises with both a sample solution and a set of tests.  The sample solution frequently did not pass all of the generated tests, however when this did happen, the test suites achieved full statement coverage in all but three cases.

\begin{figure}

\begin{tcolorbox}
\begin{Verbatim}[fontsize=\fontsize{9pt}{9pt},breaklines]
Write a function called count_rebounds that takes a list of basketball players as an input.  The function should return the total number of rebounds for the entire team.  Each element in the list is itself a list containing the player's name, their points, and their rebounds.
--Sample solution--
def count_rebounds(players):
  total = 0
  for player in players:
    total = total + player[2]
  return total
\end{Verbatim}
\end{tcolorbox}
\caption{Example output generated by Codex using the priming exercise from Figure \ref{fig:priming}.}
\label{fig:new-exercise}
\end{figure}

\begin{table}[ht!]
\centering

\caption{Analysis of generated programming exercises. 
\label{tab:dynamic_analysis_results}}
\begin{tabular}{r|cc}
\toprule
& Percentage & n out of N  \\
\midrule
Has sample solution? & 84.6\% & 203 / 240 \\
Sample solution executable? & 89.7\% & 182 / 203 \\
Has test cases? & 70.8\% & 170 / 240 \\
All tests pass? & 30.9\% & 51 / 165 \\
Full (100\%) statement coverage? & 94.1\% & 48 / 51 \\
\bottomrule
\end{tabular}
\end{table}

We also found that the vast majority of exercises (around 80\%) were entirely novel in that fragments of the problem descriptions were not indexed by any search engines.  A similar fraction of the exercises also matched the desired topics and themes.  Although this is far from perfect, there is obvious potential for generating new and useful resources in this manner and the cost of eliminating poor results (which could be automated) is almost certainly smaller than generating a large number of exercises manually.  With the addition of filtering steps that could be automated, it would be possible to generate an almost endless supply of novel resources that are contextualized to students' interests.  This also provides an interesting avenue of future work on, for example, how students use individualized, tailored exercises, and to what extent faculty will adopt AI exercise generators as part of their teaching.

\subsubsection{Code explanations}

Code explanations can be generated at different levels of abstraction, from high-level summaries to detailed explanations of every line.  We focused on the latter as these are often useful for students when debugging code.  We prompted  Codex using a simple input that consisted of the source code to be explained, the text ``Step-by-step explanation of the above program'', and finally ``1.'' to influence the output to include numbered lines.  We analysed the resulting explanations in terms of completeness and accuracy, finding that 90\% of the explanations covered all parts of the code, and nearly 70\% of the explanations for individual lines were correct.  Common errors were mostly related to relational operators and branching conditionals (e.g. where Codex stated ``less than or equal to x'' when the corresponding code was checking ``less than x'').
In recent similar work, MacNeil et al. explored a wider variety of prompt types to elicit different kinds of code explanations, and deployed these in an online interactive e-book on web software development \cite{macneil2023experiences}.  Students viewing these explanations tended to find the high-level summaries of code more useful for their learning when compared to lower-level detailed explanations of each line. Recent work has also found that LLM-created code explanations are rated more highly on average by students than code explanations created by their peers~\cite{leinonen2023comparing}.

\begin{figure}[h]
    \centering
    \includegraphics[width=.95\columnwidth]{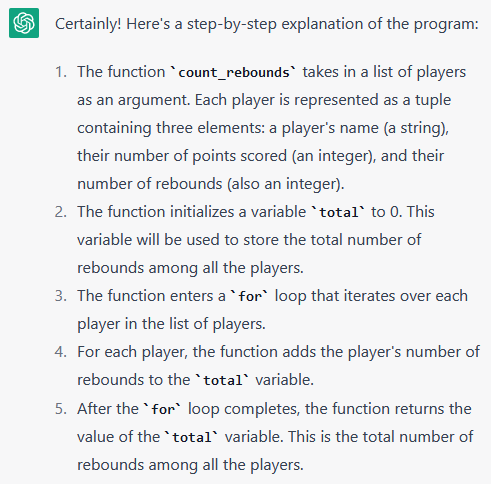}
    \caption{Explanation generated by the ChatGPT model of the code shown in the `Sample solution' area in Figure \ref{fig:new-exercise}.}
    \label{fig:chatgpt_explain}
\end{figure}

We see great potential for LLMs to be applied to the production of a variety of learning resources relevant to computing education.  We also expect the quality and accuracy of the generated resources to improve drastically over the near term based on recent improvements.  Indeed, in this section we have described the generation of code explanations using the version of the Codex model that was available in early 2022 (specifically, `code-davinci-001').  Less than a year later, code explanations generated by models such as ChatGPT are considerably better and more consistently accurate.  Figure \ref{fig:chatgpt_explain} illustrates an example of a code explanation generated by ChatGPT when provided only the code shown in the ``Sample solution'' area in Figure \ref{fig:new-exercise} and using the same prompt for a line-by-line explanation as described in this section.





\subsection{Better Programming Error Messages}
For over six decades, researchers have identified poor Programming Error Messages (PEMs) as problematic, and significant work remains in this area~\cite{Becker2019wgpaper}. PEMs can be notoriously difficult to decipher, especially for novices,
possibly contributing to the perception that programming is overly challenging~\cite{becker2021what}. 
Recent work has attempted to put error messages into more natural language by focusing on readability, 
which has been shown to improve student understanding of error messages and the number of successful code corrections ~\cite{denny2021designing}. While it is clear that increasing the readability of PEMs is helpful to novices, doing so at scale, and across languages, remains a challenge.

Leinonen et al. recently took a different approach and used LLMs to try to improve PEMs~\cite{leinonen2023using}. They collected Python error messages that had been reported as most unreadable in prior work
and generated code examples that produced these error messages. They prompted the Codex API with both the code and error message in order to generate explanations of the PEMs and actionable fixes. They found that most of the explanations created by Codex were comprehensible, and that Codex produced an output with an explanation for most inputs. This exploratory work indicates clear potential for such methods to be used in practice, though returned fixes would still need to be checked for correctness before presenting them to students. The potential to demystify PEMs in this way is an exciting opportunity only recently made possible.

\subsection{Exemplar Solutions}

Students often seek exemplar solutions when coding, either to check against their own code or to get help when struggling.  Instructors may not have the time to prepare and publish model solutions for all programming exercises that students encounter, including current and historical test and exam questions.  Thus, AI-generated solutions can be a time-efficient option~\cite{denny2022conversing}. Moreover, given the variety of solutions that code generation tools can produce from the same input prompt, they provide a powerful route to expose students to different approaches to solve a given problem. 
Thompson et al.\ argue that providing appropriate variation during programming instruction is important, as it helps learners appreciate the trade-offs of different solution approaches~\cite{thompson2006code}.  Eckerdal and Thun\'{e} make a similar argument, drawing on variation theory to state that teachers should make available resources that highlight dimensions of variation~\cite{eckerdal2005novice}.  For most non-trivial problems, code generation tools produce a variety of correct solutions that can efficiently demonstrate this variation.  The work by Finnie-Ansley et al. on the Rainfall problem discussed earlier also supports this~\cite{finnieansley2022robots}.  

Finally, by generating exemplar solutions easily the emphasis can shift from just ensuring that code is correct to also considering factors such as quality and style.  With the ability to generate syntactically correct solutions automatically, assessment can focus on the differences between multiple correct solutions, and the need to make judgments on the style and quality of solutions.  Extensive research on the benefits of peer review, including code reviews~\cite{indriasari2020review},
suggests that it is beneficial to consider multiple solutions to a problem, even if some of them are flawed. Code generation models can be used to create solutions of varying quality, and these can be used for assessment tasks that require students to apply the critical analysis skills needed for code  evaluation.
This can facilitate discussions about different approaches and the quality of solutions, and provide opportunities for refactoring exercises~\cite{finnieansley2022robots}.

\subsection{New Pedagogical Approaches}

  
\subsubsection{Algorithms early}
In a traditional CS1 course, the initial focus usually begins with syntax and basic programming principles, and it can take time for students to become proficient in these fundamentals.  If code generation models can be utilized to handle low-level implementation tasks, this may enable students to start focusing on higher-level algorithms earlier. This approach has some similarity with the use of block-based environments for novices, which abstract away the complexities of syntax and allow students to focus on algorithmic issues. As a result, teaching could initially concentrate more on algorithms and problem-solving aspects, using automatic code generation for implementation, and defer in-depth and nuanced discussions of syntax until later.

\subsubsection{Specification-focused tasks}

A similar shift could be seen away from implementation and towards specification.  Novices are typically presented with very carefully specified problems with clear and unambiguous problem statements.  Such detailed specifications provide excellent context for code generation models to generate correct code solutions.  New types of problems could task students with developing the specifications themselves, perhaps by inferring patterns from provided test suites
.  The student-written specifications could then be provided to code models for the implementation step, and iteratively refined.

\subsubsection{Explaining algorithmic concepts clearly}
It is well known that the outputs produced by large language models are very sensitive to their inputs~\cite{reynolds2021computation}.  In fact, ``prompt engineering,'' where effective prompts are crafted, has emerged as a distinct (and nascent) skill when working with these models.  For example, when using Codex to solve probability and statistics problems, engineering the prompt to include explicit hints on the strategy for solving a problem is extremely effective~\cite{tang2022solving}.  Denny et al. found that prompt engineering strategies which described algorithmic steps were effective for solving programming tasks for which Copilot initially generated solutions that were incorrect~\cite{denny2022conversing}.  Other recent work has shown that developers are more successful working with Copilot when they decompose larger programming statements into smaller tasks and then explicitly prompt Copilot for each of the subtasks~\cite{barke2022grounded, jiang2022discovering}.   It is likely that students will need to develop new skills to communicate effectively with these models.  A key skill will be the ability to describe the computational steps they wish to achieve in natural language as a way of guiding the model to produce valid outputs.



\subsubsection{A focus on refactoring}

Students sometimes experience difficulty getting started 
on programming assignments, sometimes referred to as the programmer's writer's block. Recent work found that Copilot can help students overcome this barrier by immediately providing starter code, enabling them to build upon existing code rather than starting from scratch with a blank code editor~\cite{vaithilingam2022expectation}. This approach may require a shift in focus towards tasks such as rewriting, refactoring, and debugging code, but it provides the opportunity to help students maintain momentum in a realistic setting where the ability to evaluate, rewrite, and extend code is often more important than writing every line of code from scratch.

%
\subsubsection{A focus on software testing}


Students sometimes submit programs that they do not fully understand.  Code generation tools are likely to exacerbate this issue, given the ease with which students can generate large blocks of opaque and untested code.  
Buck and Stucki argue that students should be gradually exposed to complexity, and that one must understand the fundamental building blocks before creating something larger that uses them. 
An explicit focus on testing may help students to develop an understanding of the code they are generating with tools like Copilot.  This is certainly not a new pedagogical approach but AI code generation may impart momentum in this area.  
Edwards, 
an advocate for test-driven development (TDD), asserted in 2003 that 
it can help students ``read and comprehend source code, envision how a sequence of statements will behave, and predict how a change to the code will result in a change in behavior''~\cite{edwards2003rethinking}.  However, it does not seem that computing educators 
fully embraced TDD or related pedagogies on a wide scale.

Since that time, TDD has become an important programming methodology in the field.
A major tenet of TDD is that even when the code base is small and functionality incomplete, the code that does exist runs without error. Since the team develops tests continuously, new code has to pass a greater number of tests, leading to increased confidence in the project as it proceeds.  The relative ratio of new-code-being-tested to old-code-already-tested decreases over time.  This practice may be upended by tools like Copilot where students are regenerating large sections of code and therefore invalidating previous tests.  A possible solution is \emph{incremental testing} in which students are rewarded for effort spent writing software tests throughout the development of a project.  
This regular, early and incremental feedback builds on previous test-driven development practices and may help students better understand the code that they are writing, and the code being generated for them.  We believe that large industry shifts combined with the availability of tools like Copilot necessitate a renewed focus on test-driven development, not as a luxury, but as a core competency for this new era of teaching computing.

\subsection{Designing LLM Tools}
Programmers around the world, not just novices, will be utilizing code generators in an increasing capacity moving forward. The design of these LLM tools for programming is nascent,  providing myriad research opportunities. Exploring the integration of LLMs directly into educational environments, such as auto-graders and online textbooks, will be an important area of research moving forward.  There is a need in such environments for appropriate guardrails so that generated outputs usefully support learning, without immediately revealing solutions or overwhelming novices with the complexity or quantity of feedback.  The importance of the `steerability' of such models was highlighted as part of the recent announcement of GPT-4\footnote{\href{https://openai.com/research/gpt-4}{https://openai.com/research/gpt-4}}, which included the example of a `socratic tutor' that would respond to a student's requests with probing questions rather than revealing answers directly.  Adapting the feedback generated by LLMs to maximise learning in educational environments is likely to be an important research focus in the near future.

Concrete recommendations are already beginning to emerge from very recent work in this space. First, the over-utilization of code generators by novices will generally decrease the number of errors they see. This seems like a positive experience, though it appears they are ill-equipped to deal with the errors they do see when presented with them \cite{kazemitabaar2023studying}. This means that tools must be designed to help users (of all skill levels) through the error feedback loop. Second, generating and inserting large blocks of code may be counter-productive for users at all levels. This requires users to read through code they did not write, sometimes at a more sophisticated level than they are familiar with. Novices may be intimidated by such code generation \cite{kazemitabaar2023studying} or may spend too much time reading code that does not further their goals \cite{prather2023weird}. Therefore, AI code generators should include a way for users to control the amount of code insertion and to specify how to step through a multi-part segment of generated code. Third, the fact that AI code generators are black boxes means that programmers of all skill levels may struggle to create correct mental models of how they work, which could harm their ability to fully utilize them or 
learn from their outputs. 
Explainable AI (XAI) patterns could be helpful here, such as exposing to the user a confidence value and user skill estimation above the generated code suggestion \cite{prather2023weird}. These three suggestions are only the beginning of a long line of research on how to helpfully design usable AI code generators that empower novice learners and enhance programmer productivity.

\section{Where Do We Go From Here?}

The emergence of powerful code generation models has led to speculation about the future of the computing discipline.  
In a recent CACM viewpoint article, Welsh claims they herald the ``end of programming'' and believes there is major upheaval ahead for which few are prepared, as the vast majority of classic computer science will become irrelevant~\cite{welsh2023end}.  In an even more recent article on BLOG@CACM, Meyer is equally impressed by the breakthroughs, placing them alongside the World Wide Web and object-oriented programming as a once-in-a-generation technology, but also takes a more optimistic view\footnote{\href{https://cacm.acm.org/blogs/blog-cacm/268103-what-do-chatgpt-and-ai-based-automatic-program-generation-mean-for-the-future-of-software/fulltext}{cacm.acm.org/blogs/blog-cacm/268103-what-do-chatgpt-and-ai-based-automatic-program-generation-mean-for-the-future-of-software/fulltext}}.  In fact, Meyer predicts a resurgence in the need for classic software engineering skills such as requirements analysis, formulating precise specifications, and software verification.

Much of the commentary to date has focused on the impact to professional developers and computer science as a broad discipline.  The immediate and long-term impacts on novice programmers, students, and their educators is a conversation fewer are having. 
Experts uniquely appreciate this new technology only because they already understand the underlying fundamentals.
Novices may be able to generate code by communicating natural language descriptions to the models, but natural language is ambiguous and the ability to read and comprehend code is arguably more important now than ever. 
Code literacy skills are essential in order to critically analyze what is being produced to ensure alignment between one's intentions and the generated code.  Without the skills to read, test and verify that code does what is intended, users risk becoming mere consumers of the generated content, relying on blind faith more than developed expertise. 

Although professional developers may indeed spend less time in the future  writing `low-level' code, we argue that writing code remains a valuable way for novices to learn the fundamental concepts essential for code literacy. 
Reading and writing code are complementary and related skills, and developing both has been a longstanding goal for students learning programming.  Although we do expect to see some shift in emphasis, even in introductory courses, towards modifying code generated by AI tools, the ability to edit such outputs and compose code in today's high-level languages will likely remain a fundamental skill for students of computing.  This aligns with Yellin's recent viewpoint that as programs increase in complexity, natural language becomes too imprecise an instrument with which to specify them \cite{yellin2023premature}.  At some point, editing code directly is more effective than issuing clarifying instructions in natural language.

Tools like Copilot and ChatGPT, harnessed correctly, have the potential to be valuable assistants for this learning.  We see these tools being used to explain concepts to a broad and diverse range of learners, generate exemplar code to illustrate those concepts, and to generate useful learning resources that are contextualized to the interests of individuals.   We also see new pedagogies emerging that leverage code generation tools, including explicit teaching of effective ways to communicate with the tools, tasks that focus on problem specification rather than implementation, and a renewed focus on test driven development.  These tools will also be leveraged by professional programmers to increase productivity and even their non-coding managers to better understand the code their developers are writing.

Educators need to adapt quickly as the use of these tools proliferates. Many existing assessment approaches may no longer be effective, and students must be taught to use these tools responsibly to support their learning rather than becoming over-reliant on them. There is also a growing need to teach new topics that focus on the wider societal implications of generative AI models.  These tools raise important and challenging legal, ethical, economic, and other considerations that are currently (re)shaping the society that our students will soon enter as graduates. As educators one of our most important, and challenging, jobs 
is to prepare students for such rapidly changing landscapes. 

We believe it is imperative to get ahead of the use of these tools, incorporate them into our classrooms from the very beginning, and teach students to use them responsibly via modern pedagogies. In short, we must adopt or perish. However, adoption is not a compromise or a concession; rather it is a path to a new flourishing.

\bibliographystyle{ACM-Reference-Format}
\bibliography{references.bib}

\balance

\end{document}